\documentclass[aps,prd,twocolumn,amsmath,amssymb,floatfix,nofootinbib,superscriptaddress]{revtex4-1}
\usepackage[utf8]{inputenc}
\usepackage{xcolor}
\usepackage{amsmath}
\usepackage{amssymb}
\usepackage{graphicx}
\usepackage{natbib}
\usepackage{hyperref}
\usepackage{aas_macros}
\usepackage[normalem]{ulem}
\usepackage{siunitx}

\newcommand{\ualphamap}[1][]{\ensuremath{\alpha^{#1 \mathrm{biased}}}}
\newcommand{\uthetamap}[1][]{\ensuremath{\theta^{#1 \mathrm{biased}}}}
\newcommand{\modmap}[1]{\ensuremath{#1^{\mathrm{mod}}}}
\newcommand{\nhat}[1][]{\ensuremath{\hat{\mathbf{n}}}}
\newcommand{\obsmap}[1]{\ensuremath{#1^{\mathrm{obs}}}}
\newcommand{\lenmap}[1]{\ensuremath{#1^{\mathrm{len}}}}
\newcommand{\unlmap}[1]{\ensuremath{#1^{\mathrm{prim}}}}
\newcommand{\rotmap}[1]{\ensuremath{#1^{\mathrm{rot}}}}
\newcommand{\biasmap}[1]{\ensuremath{#1^{\mathrm{biased}}}}
\newcommand{\vect}[1]{\ensuremath{\boldsymbol{#1}}}

\newcommand{\smu}{Department of Physics,
Southern Methodist University, 3215 Daniel Ave, Dallas, Texas 75275, USA}

\begin{document}

\title{Reconstructing Cosmic Polarization Rotation with ResUNet-CMB}
\author{Eric~Guzman}
\affiliation{\smu}
\author{Joel~Meyers}
\affiliation{\smu}
\date{\today}

\begin{abstract}
Cosmic polarization rotation, which may result from parity-violating new physics or the presence of primordial magnetic fields, converts $E$-mode polarization of the cosmic microwave background (CMB) into $B$-mode polarization.
Anisotropic cosmic polarization rotation leads to statistical anisotropy in CMB polarization and can be reconstructed with quadratic estimator techniques similar to those designed for gravitational lensing of the CMB.
At the sensitivity of upcoming CMB surveys, lensing-induced $B$-mode polarization will act as a limiting factor in the search for anisotropic cosmic polarization rotation, meaning that an analysis which incorporates some form of delensing will be required to improve constraints on the effect with future surveys.
In this paper we extend the ResUNet-CMB convolutional neural network to reconstruct anisotropic cosmic polarization rotation in the presence of gravitational lensing and patchy reionization, and we show that the network simultaneously reconstructs all three effects with variance that is lower than that from the standard quadratic estimator nearly matching the performance of an iterative reconstruction method.
\end{abstract}

\maketitle

\section{Introduction}\label{sec:Intro}
Observations of the cosmic microwave background (CMB) have played a key role in establishing the current standard model of cosmology.  Upcoming CMB surveys, including those with Simons Observatory~\cite{Ade:2018sbj}, BICEP Array~\cite{Hui:2018cvg}, CCAT-prime~\cite{Aravena:2019tye}, CMB-S4~\cite{Abazajian:2016yjj}, LiteBIRD~\cite{Hazumi:2019lys}, PICO~\cite{Hanany:2019lle}, and CMB-HD~\cite{Sehgal:2019ewc} will provide exquisite measurements of the temperature and polarization anisotropies across a wide range of angular scales.  These measurements will provide critical tests of the standard cosmological model and enable improved searches for novel cosmological signals.  

One of the main targets for upcoming surveys is $B$-mode polarization sourced by primordial gravitational waves~\cite{Kamionkowski:1996ks,Seljak:1996gy}.  Gravitational lensing of the CMB~\cite{Lewis:2006fu} acts a source of confusion for primordial gravitational wave searches since lensing can convert $E$-mode polarization into $B$-mode polarization~\cite{Zaldarriaga:1998ar,Lewis:2001hp}.  Delensing the CMB, the process of using an estimate of the lensing field to reverse its effects, can improve constraints on primordial gravitational waves~\cite{Knox:2002pe,Kesden:2002ku,Seljak:2003pn} and will be necessary to achieve the goals of upcoming surveys~\cite{Abazajian:2016yjj,Abazajian:2019eic,Abazajian:2020dmr}.

Cosmic polarization rotation, or cosmic birefringence, can also convert $E$-mode polarization to $B$-mode polarization.  Cosmic birefringence can be caused by parity-violating new physics~\cite{Carroll:1989vb,Harari:1992ea,Carroll:1998zi,Lue:1998mq} or by the presence of primordial magnetic fields~\cite{Kosowsky:1996yc}.  A well-motivated class of models containing parity violation is provided by axion-like fields that may be related to dark matter or dark energy, with Chern-Simons coupling to the electromagnetic field strength~\cite{Marsh:2015xka,Ferreira:2020fam}.

Isotropic cosmic birefringence, characterized by a constant rotation angle of polarization vectors across the sky, leads to parity-violating $EB$ and $TB$ correlations in the CMB.  Searches for these correlations are challenging due to a degeneracy of the effects of isotropic cosmic birefringence with calibration error of the polarization direction of the instrument~\cite{QUaD:2008ado,WMAP:2010qai,Keating:2012ge}.  This degeneracy can be mitigated by searching for the different effects of cosmic birefringence on photons propagating over different distances such as polarized foregrounds~\cite{Minami:2019ruj,Minami:2020fin} or CMB photons scattered during reionization~\cite{Sherwin:2021vgb}.  A recent analysis of data from the Planck satellite~\cite{Planck:2018lkk} has shown a $2.4\sigma$ hint of non-zero isotropic cosmic birefringence~\cite{Minami:2020odp}.

Anisotropic cosmic birefringence, where the rotation angle takes different values along different directions, leads to non-stationary CMB statistics.  The mode-coupling induced by anisotropic cosmic birefrengence is analogous to the effect of deflection caused by gravitational lensing, and this change to CMB statistics can be exploited to reconstruct a map of the rotation angle~\cite{Kamionkowski:2008fp,Yadav:2009eb,Gluscevic:2009mm} using techniques that are very similar to lensing reconstruction~\cite{Hu:2001kj}.
Searches for anisotropic cosmic birefringence with existing CMB data have placed upper limits on the angular power spectrum of the rotation angle~\cite{Gluscevic:2012me,POLARBEAR:2015ktq,Liu:2016dcg,BICEP2:2017lpa}, with the tightest constraints coming from ACTPol~\cite{Namikawa:2020ffr} and SPTpol~\cite{SPT:2020cxx} $\ell(\ell+1)C_\ell^{\alpha\alpha}/(2\pi) < 0.10\times10^{-4}~\mathrm{rad}^2\, (0.033~\mathrm{deg}^2, \, 95\%~\mathrm{CL})$.
Future CMB surveys are expected to dramatically improve constraints on cosmic birefringence~\cite{Pogosian:2019jbt}.

Much like the search for primordial gravitational waves, $B$-mode polarization induced by CMB lensing acts as a source of confusion in searches for cosmic polarization rotation~\cite{Yadav:2009eb}.  Therefore, delensing the CMB is necessary in order to achieve the tightest constraints on cosmic birefringence.  Delensing also allows for tighter cosmological parameter constraints~\cite{Green:2016cjr}, reduces lensing-induced covariance, aids the search for primordial non-Gaussianity~\cite{Coulton:2019odk}, and reduces the variance in the reconstruction of other CMB secondaries including patchy reionization~\cite{Su:2011ff}.

CMB delensing using an estimate of the lensing field reconstructed from the observed CMB map will provide the best delensing performance at the sensitivity of upcoming surveys~\cite{Smith:2010gu}.  The process of internal delensing impacts the statistics of the delensed CMB in a way that can lead to biases if not treated carefully~\cite{Sehgal:2016eag,BaleatoLizancos:2020mic}.  Internal delensing has been demonstrated on real data~\cite{Carron:2017vfg,POLARBEAR:2019snn,SPTpol:2020rqg}, and improved delensing techniques are being developed and refined~\cite{Hirata:2002jy,Hirata:2003ka,Carron:2017mqf,Millea:2017fyd,Carron:2018lcr,Millea:2020cpw,Diego-Palazuelos:2020lme}.

Machine learning has been shown to hold promise for CMB lensing reconstruction and delensing~\cite{Caldeira:2018ojb}.  In a previous paper, we extended the network described in Ref.~\cite{Caldeira:2018ojb} and showed that the resulting network, ResUNet-CMB, performs very well at simultaneously delensing and reconstructing the modulation resulting from patchy reionization~\cite{Guzman:2021nfk}.  We showed that ResUNet-CMB achieved nearly optimal performance on patchy reionization reconstruction over a wide range of angular scales.

In this paper, we demonstrate that ResUNet-CMB is capable of achieving excellent performance when applied to the reconstruction of cosmic polarization rotation as well.  In particular, we show that ResUNet-CMB reconstructs anisotropic cosmic birefringence with a variance that is lower than could be achieved by applying the standard quadratic estimator to the lensed and rotated CMB.  Achieving this high fidelity reconstruction of cosmic polarization rotation required very little modification of the network architecture of ResUNet-CMB.  It is very encouraging that ResUNet-CMB successfully and straightforwardly generalizes to incorporate cosmic birefringence reconstruction.  The physical effects of gravitational lensing, patchy reionization, and cosmic polarization rotation are quite different, and still the machine learning architecture is able to handle them all. 

This paper is organized as follows.  In Sec.~\ref{sec:QE}, we review the quadratic estimator for fields that produce statistical anisotropy of the CMB and discuss iterative reconstruction techniques.  Section~\ref{sec:DeepLearning} describes the ResUNet-CMB architecture, simulation data pipeline, and training methods.  We present the predictions of ResUNet-CMB and draw comparisons with the quadratic estimator in Sec.~\ref{sec:Results}. We conclude in Sec.~\ref{sec:Conclusion}.

\section{Quadratic Estimator}\label{sec:QE}
The standard technique for reconstruction of the CMB lensing potential makes use of a quadratic estimator~\cite{Hu:2001kj}.  A formalism similar to the lensing quadratic estimator can also be applied to reconstruct other distortion fields such as patchy reionization~\cite{Dvorkin:2008tf,Dvorkin:2009ah} and cosmic birefringence~\cite{Kamionkowski:2008fp,Yadav:2009eb,Gluscevic:2009mm}. In this section we will discuss the quadratic estimator and how reconstruction of the anisotropic rotation angle is impacted by the effects of lensing and patchy reionization.

Weak gravitational lensing of the CMB occurs when the photons traveling towards our telescopes are gravitationally deflected by cosmological structure along their path. The photons we receive appear to originate from a position that differs by a small angle from their true origin. In real space, this deflection alters the $T$, $Q$, and $U$ maps of the CMB as
\begin{align}
    \label{eqn:LensedFields}
    \lenmap{T}(\nhat) &= \unlmap{T}(\nhat + \nabla \phi (\nhat)) \, , \nonumber\\
    (\lenmap{Q} \pm i\lenmap{U})(\nhat) &= (\unlmap{Q} \pm i\unlmap{U})(\nhat + \nabla \phi (\nhat)) \, ,
\end{align}
where $\nhat$ is the line-of-sight direction, $\phi$ is the CMB lensing potential, the `prim' superscript identifies the primoridial CMB maps, and the `len' superscript denotes the lensed maps.
The deflection of fluctuations leads to non-stationary statistics of observed maps giving off-diagonal mode coupling in harmonic space.
The induced mode coupling can be used to reconstruct a map of the lensing potential~\cite{Hu:2001kj}.

For a field $\theta$ that distorts CMB maps, the off-diagonal mode coupling is of the form 
\begin{align}
    \label{eqn:GeneralModeCoupling}
    \langle X(\vect{\ell}_1) Y(\vect{\ell}_2) \rangle_\mathrm{CMB} = f_{XY}^\theta(\vect{\ell}_1,\vect{\ell}_2)\theta(\vect{\ell}) \, ,
\end{align}
for $\vect{\ell}_1 \neq \vect{\ell}_2$ and $\vect{\ell}=\vect{\ell}_1 + \vect{\ell}_2$, where $X$ and $Y$ are CMB temperature or polarization fluctuations $X,Y\in\{T,E,B\}$.  The general form of the quadratic estimator for any such distortion field can be written as
\begin{align}
    \label{eqn:GeneralQE}
    \hat{\theta}_{XY}(\vect{\ell}) = N_{X Y}^{\theta}(\ell) \int{\frac{d^2 \vect{\ell}_{1}}{(2 \pi)^2 } \obsmap{X}(\vect{\ell}_{1}) \obsmap{Y}(\vect{\ell}_{2}) F^{\theta}_{X Y}(\vect{\ell}_{1} , \vect{\ell}_{2})} \, ,
\end{align}
where the superscript `obs' refers to the observed maps including instrumental noise, $N_{XY}^\theta(\ell)$ is a normalization constant chosen to make the estimator unbiased $\langle \hat{\theta}_{XY}(\vect{\ell}) \rangle_{\mathrm{CMB}} = \theta(\vect{\ell})$, and $F_{XY}^\theta$ is a filter (often chosen to minimize the variance of the estimator).  The CMB noise is characterized by power spectra given here by
 \begin{align}
     C^{T T, \mathrm{noise}}_{\ell} &= \Delta_{T}^{2} e^{\ell^{2} \theta_{\mathrm{FWHM}}^{2} / (8 \ln{2})} \, ,\nonumber\\
     C^{E E, \mathrm{noise}}_{\ell} &= C^{B B, \mathrm{noise}}_{\ell} = \Delta_{P}^{2} e^{\ell^{2} \theta_{\mathrm{FWHM}}^{2} / (8 \ln{2})} \, .
 \end{align}
We will focus on the $EB$ quadratic estimator for which the minimum variance filter is
\begin{align}
    \label{eqn:GeneralMinVarEBFilter}
    F^{\theta}_{EB}(\vect{\ell}_{1}  , \vect{\ell}_{2}) = 
    \frac{f_{EB}^\theta(\vect{\ell}_1,\vect{\ell}_2)}
    { C_{\ell_1}^{EE,\mathrm{obs}} C_{\ell_2}^{BB,\mathrm{obs}} } \, ,
\end{align}
The normalization for the minimum variance $EB$ estimator is
\begin{align}
    \label{eqn:GeneralMinVarEBNorm}
    N^{\theta}_{E B}(\vect{\ell}) = \left[\int{\frac{d^2 \vect{\ell}_{1}}{(2 \pi)^2 } f^{\theta}_{E B}(\vect{\ell}_{1}, \vect{\ell}_{2}) F^{\theta}_{E B}(\vect{\ell}_{1}, \vect{\ell}_{2})}\right]^{-1} \, .
\end{align}
For the minimum variance filter, the normalization also gives the noise power of the reconstructed map
\begin{align}
    \label{eqn:GeneralReconNoise}
    \left\langle \hat{\theta}_{EB}(\vect{\ell}) \hat{\theta}_{EB}(\vect{\ell}') \right\rangle = (2\pi)^2 \delta(\vect{\ell} + \vect{\ell}') \left(C_\ell^{\theta\theta} + N_{EB}^\theta(\ell) \right) \, .
\end{align}

The $EB$ mode coupling induced by gravitational lensing deflection is
\begin{align}
    f^{\phi}_{E B}(\vect{\ell}_{1}, \vect{\ell}_{2}) = &\left[C_{\ell_{1}}^{E E} (\vect{\ell}_1\cdot\vect{\ell}) - C_{\ell_{2}}^{B B} (\vect{\ell}_2\cdot\vect{\ell})\right] \nonumber \\
    &\times \sin{2(\varphi(\vect{\ell}_1) - \varphi(\vect{\ell}_2))} \, .
    \label{eqn:LensingModeCoupling}
\end{align}
Cosmic polarization rotation leaves the temperature unchanged but mixes $Q$ and $U$ polarization
\begin{align}
     \label{eqn:AlphaRotation}
     \rotmap{T} &= \unlmap{T} \nonumber\\
     \rotmap{Q}(\nhat) &= \unlmap{Q}(\nhat) \cos{2 \alpha(\nhat)} - \unlmap{U}(\nhat) \sin{2 \alpha(\nhat)} \nonumber\\
     \rotmap{U}(\nhat) &= \unlmap{Q}(\nhat) \sin{2 \alpha(\nhat)} + \unlmap{U}(\nhat) \cos{2 \alpha(\nhat)} \, ,
\end{align}
which leads to an $EB$ mode coupling of the form
\begin{equation}
    \label{eqn:AlphaModeCoupling}
    f^{\alpha}_{E B}(\vect{\ell}_{1}, \vect{\ell}_{2}) = 2\left[C_{\ell_{1}}^{E E} - C_{\ell_{2}}^{B B}\right] \cos{2(\varphi(\vect{\ell}_1) - \varphi(\vect{\ell}_2))} \, .
\end{equation}
Inhomogeneous reionization leads to a CMB optical depth that differs across the sky $\tau(\nhat)$, thereby causing a modulation of CMB fluctuations~\cite{Hu:1999vq,Santos:2003jb,Zahn:2005fn,McQuinn:2005ce,Dore:2007bz,Dvorkin:2008tf,Dvorkin:2009ah,Battaglia:2012im,Park:2013mv,Alvarez:2015xzu,Paul:2020fio,Choudhury:2020kzh}
\begin{align}
     \label{eqn:ModulatedFields}
     \modmap{T}(\nhat) &=  \unlmap{T}(\nhat) e^{-\tau(\nhat)}  \nonumber\\
     (\modmap{Q} \pm i\modmap{U})(\nhat)  &=  (\unlmap{Q} \pm i\unlmap{U})(\nhat) e^{-\tau(\nhat)} \, .
\end{align}
Patchy reionization also generates new polarization fluctuations through the scattering of remote temperature quadrupole anisotropies~\cite{Dvorkin:2008tf,Dvorkin:2009ah}, though we will neglect these scattering contributions here.  The $EB$ mode coupling from modulation is given by
\begin{equation}
     \label{eqn:PatchyReionizationModeCoupling}
     f^{\tau}_{E B}(\vect{\ell}_{1}, \vect{\ell}_{2}) = \left[C_{\ell_{1}}^{E E} - C_{\ell_{2}}^{B B}\right] \sin{2(\varphi(\vect{\ell}_1) - \varphi(\vect{\ell}_2))} \, .
 \end{equation}

Primordial $B$-mode polarization is not generated by density fluctuations at first order in perturbation theory, and the contribution from primordial gravitational waves is observationally constrained to be small~\cite{BICEP2:2018kqh}.  Distortions of the CMB, such as deflection by gravitational lensing, cosmic polarization rotation, and modulation due to patchy reionization, convert $E$-mode polarization to $B$-mode polarization.  As can be seen from Eqs.~\eqref{eqn:GeneralMinVarEBFilter} and \eqref{eqn:GeneralMinVarEBNorm}, the variance of the reconstruction of any distortion with the $EB$ estimator depends upon the total observed $B$-mode power, including the secondary $B$-mode polarization produced by distortion fields.  Lensing-induced $B$ modes make the dominant contribution to the observed $B$-mode power in the absence of noise and foregrounds on the scales of interest.

The variance of reconstructions of distortion fields can thereby be improved if we can reverse the effects of all such distortions, and in particular if we can remove the additional $B$-mode power that these effects cause.  For example, the $B$-mode power produced by gravitational lensing of $E$ modes (to leading order in the gradient expansion) is
\begin{align}
    C_{\ell}^{BB,\mathrm{len}} =& \int{\frac{d^2 \vect{\ell}_{1}}{(2 \pi)^2 }} \left[ (\vect{\ell}_1 \cdot \vect{\ell}_2) \sin{2(\varphi(\vect{\ell}_1) - \varphi(\vect{\ell}))} \right]^2 \nonumber \\
    & \times C_{\ell_2}^{\phi\phi} C_{\ell_1}^{EE}  \, .
    \label{eq:LensedBB}
\end{align}
The $B$-mode power can be reduced by removing an estimate of the $B$ modes constructed from the measurement of the $E$-mode polarization and the reconstructed lensing potential
\begin{align}
    C_{\ell}^{BB,\mathrm{len},\mathrm{res}} =& \int{\frac{d^2 \vect{\ell}_{1}}{(2 \pi)^2 }} \left[ (\vect{\ell}_1 \cdot \vect{\ell}_2) \sin{2(\varphi(\vect{\ell}_1) - \varphi(\vect{\ell}))} \right]^2 \nonumber \\
    & \times C_{\ell_2}^{\phi\phi} C_{\ell_1}^{EE} \left[ 1 - \frac{C_{\ell_2}^{\phi\phi}}{C_{\ell_2}^{\phi\phi,\mathrm{obs}}} \frac{C_{\ell_1}^{EE}}{C_{\ell_1}^{EE,\mathrm{obs}}} \right] \, ,
    \label{eq:LensingResidualBB}
\end{align}
where $C_{\ell}^{\phi\phi,\mathrm{obs}} = C_{\ell}^{\phi\phi}+N^{\phi}(\ell)$.
Using the reduced $B$-mode power, one can obtain a lower variance estimate of the lensing field with the $EB$ estimator, which can subsequently be used to remove a larger fraction of the lensed $B$ modes~\cite{Smith:2010gu}.  Iterating this procedure to convergence gives a variance on the reconstructed lensing field which matches closely with the maximum likelihood reconstruction~\cite{Smith:2010gu,Hirata:2003ka}.

A similar procedure can be followed to demodulate the CMB, giving
\begin{align}
    C_{\ell}^{BB,\mathrm{mod},\mathrm{res}} =& \int{\frac{d^2 \vect{\ell}_{1}}{(2 \pi)^2 }} \left[  \sin{2(\varphi(\vect{\ell}_1) - \varphi(\vect{\ell}))} \right]^2 \nonumber \\
    & \times C_{\ell_2}^{\tau\tau} C_{\ell_1}^{EE} \left[ 1 - \frac{C_{\ell_2}^{\tau\tau}}{C_{\ell_2}^{\tau\tau,\mathrm{obs}}} \frac{C_{\ell_1}^{EE}}{C_{\ell_1}^{EE,\mathrm{obs}}} \right] \, ,
    \label{eq:ModulationResidualBB}
\end{align}
and also to derotate the CMB
\begin{align}
        C_{\ell}^{BB,\mathrm{rot},\mathrm{res}} =& \int{\frac{d^2 \vect{\ell}_{1}}{(2 \pi)^2 }} \left[  \cos{2(\varphi(\vect{\ell}_1) - \varphi(\vect{\ell}))} \right]^2 \nonumber \\
    & \times 4 C_{\ell_2}^{\alpha\alpha} C_{\ell_1}^{EE} \left[ 1 - \frac{C_{\ell_2}^{\alpha\alpha}}{C_{\ell_2}^{\alpha\alpha,\mathrm{obs}}} \frac{C_{\ell_1}^{EE}}{C_{\ell_1}^{EE,\mathrm{obs}}} \right] \, .
    \label{eqRotationResidualBB}
\end{align}
If we treat the effects of lensing, modulation, and rotation as entirely independent, the observed $B$-mode power spectrum takes the form $C_\ell^{BB,\mathrm{obs}} \simeq C_\ell^{BB,\mathrm{prim}} + C_\ell^{BB,\mathrm{len}} + C_\ell^{BB,\mathrm{mod}} + C_\ell^{BB,\mathrm{rot}} + C_\ell^{BB,\mathrm{noise}}$.  The presence of lensing-induced $B$-mode power increases the variance of the modulation and rotation reconstruction, and similarly for each contribution to the total $B$-mode spectrum.  We can improve the reconstruction of each field by delensing, demodulating, and derotating to remove $B$-mode power.  This whole procedure of lensing reconstruction, delensing, modulation reconstruction, demodulation, polarization rotation reconstruction, and derotation can be iterated to convergence to estimate the maximum likelihood reconstruction noise in the presence of all three effects.

This iterative procedure is only an approximation, since it fails to account for the fact that the effects of lensing, modulation, and rotation are not independent.  In the simulations we employ here, the polarization maps are first anisotropically rotated by $\alpha$, then modulated by $\tau$, and finally lensed by $\phi$.  The effect of this series of deformations is not fully captured by a simple sum of the $B$-mode power from each effect applied in isolation.  Furthermore, the real physical effects are not so distinctly separated in time, though we treat them as such for simplicity in the simulations we present.

Implementing an iterative reconstruction scheme at map level is a non-trivial task, though iterative lensing reconstruction schemes have been developed~\cite{Hirata:2002jy,Seljak:2003pn,Carron:2017vfg,Millea:2017fyd,Millea:2020cpw}, and  internal delensing of CMB maps has been demonstrated on data from Planck~\cite{Carron:2017vfg,Planck:2018lbu}, the Atacama Cosmology Telescope~\cite{ACT:2020goa}, and the South Pole Telescope~\cite{Millea:2020iuw}.
Map-level iterative reconstruction of multiple sources of statistical anisotropy poses additional challenges, since one must deal at each step with biases that may arise.  For example, the presence of lensing leads to a bias in the reconstruction of patchy reionization maps~\cite{Su:2011ff} and cosmic polarization rotation spectra~\cite{Namikawa:2016fcz,SPT:2020cxx}.  For the purpose of this work, we do not attempt to construct a map-level iterative reconstruction scheme based on the quadratic estimator.  Instead, we use the scheme described above at the level of the power spectra to estimate the reconstruction noise that would be expected assuming the existence of a map-level iterative reconstruction procedure that was able to deal with these complications without a significant impact on the resulting reconstruction noise.
In Sec.~\ref{sec:Results}, we compare the noise of the cosmic polarization rotation reconstruction derived from this idealized iterative procedure to that obtained from ResUNet-CMB.

\begin{figure*}[t!]
    \centering
    \includegraphics[width=0.95\textwidth]{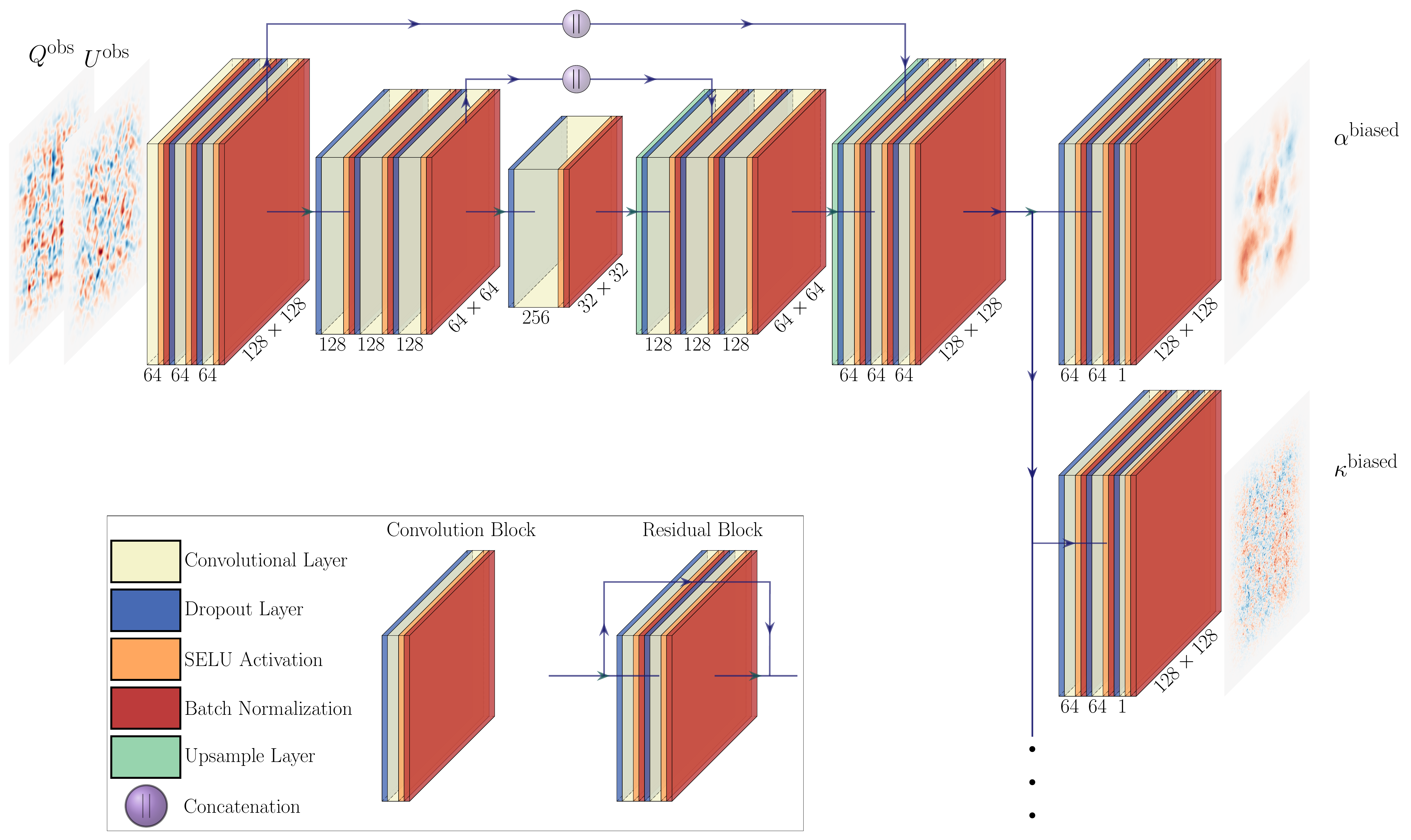}
    \caption{
    ResUNet-CMB architecture showing two of four output branches and with all residual connections excluded for visual clarity. 
    The input layer consists of the ($\obsmap{Q}$, $\obsmap{U}$) CMB polarization maps concatenated along the color channel.
    The maps ($\biasmap{\alpha}$, $\biasmap{\kappa}$, $\biasmap{\tau}$, $\biasmap{E}$) are the output of the final batch normalization layers of their respective branches. 
    The numbers appearing on the bottom of each convolutional layer represent the number of filters that layer contains and the image size is displayed on side of every third batch normalization layer. 
    In the encoder phase down-sampling is done by changing the stride of convolution layers to a value of 2. 
    (The graphic was made with publicly available code from \url{https://github.com/HarisIqbal88/PlotNeuralNet}.)}
    \label{fig:ResUNet-Structure}
\end{figure*}

\section{ResUNet-CMB Architecture and Methods}\label{sec:DeepLearning}

Convolutional neural networks (CNNs) are a type of deep learning network often used for computer vision tasks. Since the data obtained from many cosmological observations can be processed as two- and three-dimensional images, CNNs are well-suited to several aspects of cosmological simulations and data analysis. For example, some current CNN applications in cosmology include producing full-sky CMB simulations~\cite{Han:2021unz}, identification of HII regions in reionization~\cite{Bianco:2021abd}, analysis of dark matter substructure~\cite{Alexander:2020mbx, Vattis:2020kaa}, and cosmic velocity field reconstruction~\cite{Wu:2021jsy}.

In this paper we use a particular type of CNN, a ResUNet~\cite{kayalibay2017cnnbased, milletari2016vnet, Zhang_2018}, to reconstruct anisotropic cosmic birefringence imprinted in the polarization of simulated CMB maps in the presence of both lensing and patchy reionization effects. As shown in Refs.~\cite{Caldeira:2018ojb, Guzman:2021nfk}, ResUNets are capable of detecting and recovering the distortions in simulated CMB maps at high fidelity.

In this section we will review the design and training of the ResUNet-CMB architecture introduced in Ref.~\cite{Guzman:2021nfk}, noting the modifications we made to the network and data pipeline in order to apply the framework to the new task of reconstructing cosmic birefringence.

\subsection{ResUNet-CMB}

We start from the ResUNet-CMB network described in Ref.~\cite{Guzman:2021nfk} which is a modified version of the ResUNet from Ref.~\cite{Caldeira:2018ojb}.  We extend the architecture in order to simultaneously reconstruct anisotropic cosmic polarization rotation $\alpha$, the lensing convergence $\kappa$, the patchy reionization modulation $\tau$, and the primordial $E$-mode polarization $\unlmap{E}$. The ResUNet-CMB architecture\footnote{The code for the updated ResUNet-CMB architecture and the new data pipeline is located at \url{https://github.com/EEmGuzman/resunet-cmb}.} is shown in Fig.~\ref{fig:ResUNet-Structure}. The network is implemented using the Keras package~\cite{chollet2015keras} of Tensorflow 2.0~\cite{tensorflow2015-whitepaper}.

The inputs to the network are simulated maps of the rotated, modulated, and lensed polarization plus noise described with Stokes parameters ($\obsmap{Q}$, $\obsmap{U}$) concatenated along the color channel.
The network has four output maps, ($\biasmap{\alpha}$, $\biasmap{\kappa}$, $\biasmap{\tau}$, $\biasmap{E}$), each with its own branch.
The superscripts `biased' indicate that the outputs of the network exhibit a multiplicative bias, the correction of which will be described in Sec.~\ref{sec:Results}.

Only two outputs are shown in Fig.~\ref{fig:ResUNet-Structure} for clarity, and the other output branches have an identical structure.
For the remainder of this paper we will refer to the network used in Ref.~\cite{Guzman:2021nfk} as the 3-output network and that in the current paper as the 4-output network.
The 4-output network has a total of \num{5494028} trainable parameters, \num{206915} more than the 3-output network.

The primary building block of the network is the convolution block shown in the legend of Fig.~\ref{fig:ResUNet-Structure}. 
Each convolution block, apart from the first and last blocks in the network, consists of a dropout layer with a drop value of 0.3, a convolutional layer using `same' padding, an activation layer using the scaled exponential linear unit (SELU) function~\cite{klambauer2017selfnormalizing}, and a batch normalization layer in that order.
The first convolution block does not contain a dropout layer in order to prevent the irreversible loss of initial information.
In the final convolution block of each output branch, the activation layer is set to a linear activation function instead of SELU. 

Accuracy of the predictions from CNNs can be negatively affected with increasing network depth~\cite{he2014convolutional,he2015deep,srivastava2015highway}. Residual connections, which can help mitigate this problem, take the input of a convolution block and add it element-wise to the output of a different convolution block~\cite{he2015deep,he2016identity,balduzzi2018shattered}. In the ResUNet-CMB architecture, the input of a convolution block is added element-wise to the output of the subsequent block. The residual connection over two convolution blocks is called a residual block and is shown in the legend of Fig.~\ref{fig:ResUNet-Structure}. The first residual connection of the network is from the concatenated (\obsmap{Q}, \obsmap{U}) layer to the input of the third convolution block. The next residual connection is from the input of the third convolution block to that of the fifth convolution block. This pattern of a connection every two blocks continues with the exception of the final convolution block in each output branch where no residual connection is placed. Like the architecture from Ref.~\cite{Guzman:2021nfk}, we include a batch normalization layer in each residual connection, placed after a convolutional layer if one is present, since we found this additional layer reduced validation loss across all outputs.

The network includes skip connections in addition to residual connections. The skip connections are concatenations along the color channel of the output of convolution blocks in the encoder phase of the network with the input of convolution blocks of the same size in the decoder phase of the network~\cite{ronneberger2015unet}. The skip connections help mitigate some localization information lost during down-sampling by providing high resolution information to convolution blocks in the decoder phase.

\subsection{Data pipeline and network training}
\label{sec:DataPipeNetTraining}

We generate maps using the same set of cosmological parameters as in Ref.~\cite{Guzman:2021nfk} in order to ease comparisons among the results.  The fiducial cosmology is defined by: $H_{0} = 67.9$~km~s$^{-1}$~Mpc$^{-1}$, $\Omega_{b}h^{2} = 0.0222$, $\Omega_{c}h^{2} = 0.118$, $n_{s} = 0.962$, $\tau = 0.0943$, and $A_{s} = 2.21\times 10^{-9}$.
The primordial CMB and lensing power spectra resulting from the chosen cosmology are produced with the software \texttt{CAMB}\footnote{\url{https://camb.info}}~\cite{Lewis:1999bs}. For the patchy reionization spectrum, $C_\ell^{\tau\tau}$, we use the $\bar{\tau} = 0.058$ model of Ref.~\cite{Roy:2018gcv}. We include anisotropic cosmic birefringence described by a scale-invariant power spectrum of the form $\ell^2 C_\ell^{\alpha\alpha}/(2\pi) = 0.014$~deg$^2$, which is about a factor of 2 smaller than the current $95\%$ upper limit from ACTPol~\cite{Namikawa:2020ffr} and SPTpol~\cite{SPT:2020cxx}.

We simulate two-dimensional maps that cover a $5^{\circ} \times 5^{\circ}$ patch of sky with $128 \times 128$ pixels by using a modified version of \texttt{Orphics}\footnote{\url{https://github.com/msyriac/orphics}}. In total eight different types of maps are generated, ($\unlmap{Q}$, $\unlmap{U}$, $\unlmap{E}$, $\alpha$, $\tau$, $\kappa$, $\obsmap{Q}$, $\obsmap{U}$).
To obtain ($\obsmap{Q}$, $\obsmap{U}$) we first take the maps ($\unlmap{Q}$, $\unlmap{U}$) and rotate them using $\alpha$ according to Eq.~\eqref{eqn:AlphaRotation}. Next, the ($\rotmap{Q}$, $\rotmap{U}$) maps are modulated by $\tau$ then lensed by $\kappa$. 
Lastly, we smooth the maps with a Gaussian beam, add a noise map, and apply a 1.5$^{\circ}$ cosine taper to the result.

The procedure to create the training, validation, test, and prediction data sets as well as the details of the pre-processing of the training data and post-processing of the predictions follows closely the discussion of Ref.~\cite{Guzman:2021nfk} modified only by the definition of the null maps and by including the additional output for $\alpha$. We restate the key points here.

We initially generate \num{70000} sets of the 8 maps described above
for four different experimental configurations.
We simulate the same set of experiments as in Ref.~\cite{Guzman:2021nfk}: a noiseless experiment; one with $\Delta_T=0.2$~$\mu$K-arcmin and $\theta_\mathrm{FWHM} = 1.0'$; and experiments with $\Delta_T=1$~$\mu$K-arcmin and 2~$\mu$K-arcmin each with $\theta_\mathrm{FWHM} = 1.4'$.
For each experiment, $\Delta_P = \sqrt{2}\Delta_T$.
Note that due to the size and resolution of our maps, the beam scale is smaller than our pixel scale and plays little role.  However, we make this choice to match the design of future experiments like CMB-S4~\cite{Abazajian:2019eic} and to ease comparisons with previous work~\cite{Su:2011ff,Caldeira:2018ojb,Guzman:2021nfk}.
Of the \num{70000} sets of maps generated, for each noise level, 20\% are randomly selected to be null maps. We define the null maps to be ($Q$, $U$) maps that are not rotated by $\alpha$
but are modulated by $\tau$ then lensed by $\kappa$.
Note that this is different than the definition of null maps used in Ref.~\cite{Guzman:2021nfk}, where in that case both $\tau$ and $\kappa$ were set to zero in the null maps.

The \num{70000} simulated maps are then split into training, validation, and test sets with a ratio of 80:10:10, respectively. In addition to the above three data sets, we produce an additional set of \num{7000} simulations called the prediction set which does not include null maps. Most of the analyses reported in Sec.~\ref{sec:Results} are derived from predictions made on the prediction set.

The training process also follows that of Ref.~\cite{Guzman:2021nfk}. Training was done on a single Nvidia Tesla V100 32GB GPU with a batch size of 32. We use the Adam optimizer~\cite{kingma2014adam} set with default parameters and choose a mean squared error loss function. An initial learning rate of 0.25 is used with a 50\% reduction in its value after three epochs of no improvement in total validation loss. Training is stopped if there is no decrease in total validation loss after 10 consecutive epochs, and the model with the lowest loss is saved. In total, four networks are trained -- one for each experiment.

\begin{figure*}[t!]
    \centering
    \includegraphics[width=\textwidth]{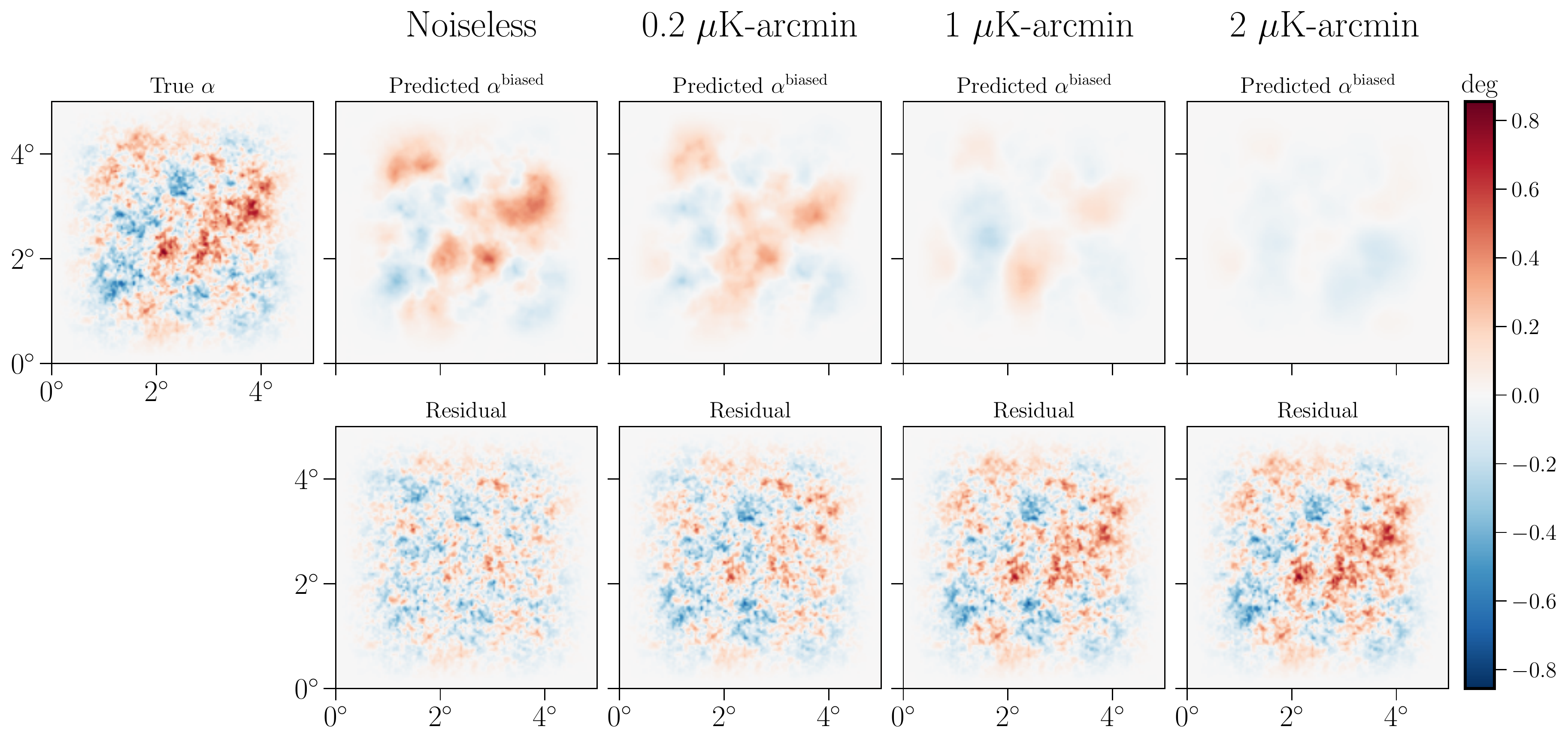}
    \caption{
    Example ResUNet-CMB predictions of anisotropic cosmic polarization rotation $\biasmap{\alpha}$ from fully trained networks for each of four experimental configurations along with the noiseless true $\alpha$ map. Residual maps are computed as the truth minus the prediction. 
    }
    \label{fig:map_results_2noiselvls}
\end{figure*}
%
Network training time for each experiment varies due to the random initialization of the learning process. On average, the noiseless, 0.2~$\mu$K-arcmin, and 1~$\mu$K-arcmin networks each take approximately 15 hours to train and converge after about 60 epochs while the 2~$\mu$K-arcmin network takes only 7 hours and about 30 epochs on average to finish training.

While the training in this paper was done on a V100 GPU, we found using a Nvidia Tesla P100 16GB GPU with a batch size of 16 was also viable. Training with a batch size of 16 increased training time by a few hours, but the VRAM usage peaked at approximately 13.2 GB for both 3- and 4-output ResUNet-CMB models.
Once the model is trained, generating a single set of predictions of the maps, ($\alpha$, $\kappa$, $\tau$, $E$), on the V100 GPU with a batch size of 32 in the prediction function takes 0.005 seconds. 

\section{Results}
\label{sec:Results}

In this section we present results for the reconstructed maps, spectra, and noise curves for predictions made from the 4-output architecture and assess the performance of ResUNet-CMB.
We compare the results to the predictions of the quadratic estimator and also to the 3-output network described in Ref.~\cite{Guzman:2021nfk}.

As previously described in Refs.~\cite{Guzman:2021nfk} and \cite{Caldeira:2018ojb}, all of the output maps generated by the network are biased.  For the rest of this paper we will refer to the outputs of the network prior to bias correction as ($\biasmap{\alpha}$, $\biasmap{\kappa}$, $\biasmap{\tau}$, $\biasmap{E}$), and the predicted maps after applying the bias correction described below will be labeled with a hat, e.g.~$\hat{\alpha}$.

\subsection{Reconstructed signal}

First we examine the $\alpha$ maps reconstructed by ResUNet-CMB.
Figure~\ref{fig:map_results_2noiselvls} shows the map-level $\biasmap{\alpha}$ predictions compared to the noiseless truth map for all four experiments. The bottom row of residual maps in the figure are defined as the truth minus the predictions. The large scale features appear to be faithfully reconstructed at all four noise levels, while only the lowest noise levels capture structure on small scales. As expected, as the instrument noise increases, the reconstruction worsens leaving larger residuals. The residual power spectrum, which we define as the power spectrum of the residual maps in Fig.~\ref{fig:map_results_2noiselvls}, is an average of 14\% larger over the range of $56<\ell<560$ for the 1~$\mu$K-arcmin experiment than in the noiseless case.

Map-level results for the remaining three outputs, ($\biasmap{\kappa}$, $\biasmap{\tau}$, $\biasmap{E}$), are visually very similar to those found in Ref.~\cite{Guzman:2021nfk} despite the additional source of statistical anisotropy. A more quantitative comparison of the predictions of the 4-output network and the 3-output network will be described in Sec.~\ref{subsec:3to4}.

\begin{figure}[ht!]
    \centering
    \includegraphics[width=\linewidth]{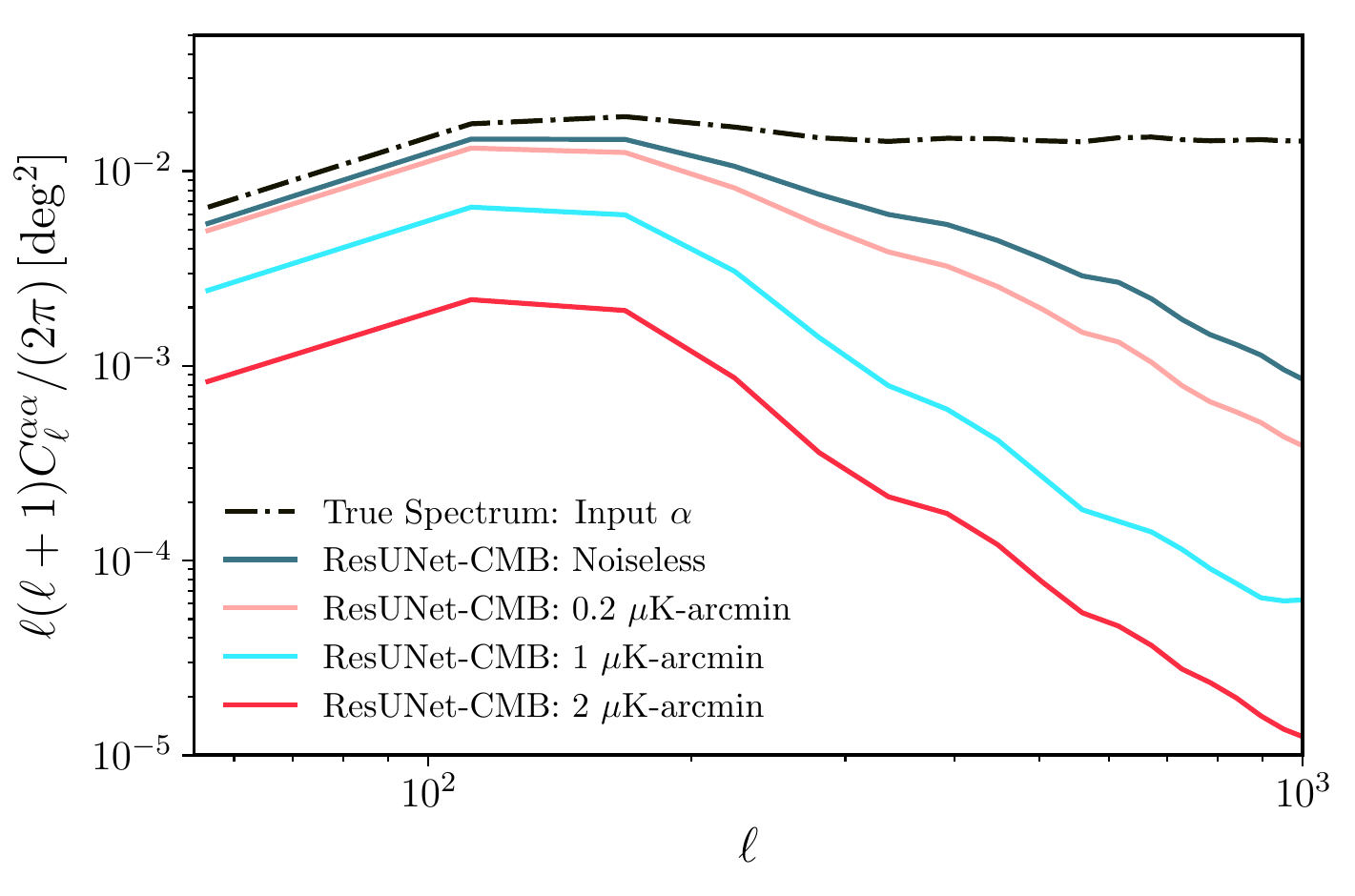}
    \caption{
    Power spectra of $\biasmap{\alpha}$ prediction maps from the ResUNet-CMB network averaged over the \num{7000} simulated CMB maps from the prediction data set for each of the four expriments (solid lines) plotted against the average power spectrum of the true $\alpha$ maps (black dashed-dot).
    }
    \label{fig:alpha_ps}
\end{figure}

Next, we present the reconstructed average pseudo-power spectra for $\alpha$, $\langle C_{\ell}^{\biasmap{\alpha} \biasmap{\alpha}} \rangle$, in Fig.~\ref{fig:alpha_ps}. To calculate $\langle C_{\ell}^{\biasmap{\alpha} \biasmap{\alpha}} \rangle$ for each noise level, we take the $\alpha$ predictions made on the prediction data set and find the power spectrum of each individual $\biasmap{\alpha}$ map. We then average over all \num{7000} spectra and divide the result by the mean of the squared taper applied to each map to get a rough window-corrected average spectrum. The true spectrum shown in Fig.~\ref{fig:alpha_ps} and below, is calculated by finding the power spectrum of the truth $\alpha$ maps in the prediction data set, averaging the result, and applying the same window correction.
The peak of the reconstructed power is in the range $56 < \ell < 170$ for all four noise levels. At $\ell>170$, the reconstructed power decreases with increasing $\ell$.
As with the map-level results, the other outputs ($\biasmap{\kappa}$, $\biasmap{\tau}$, $\biasmap{E}$) show reconstruction power similar to those from Ref.~\cite{Guzman:2021nfk}.

\begin{figure*}[t!]
    \centering
    \includegraphics[width=\textwidth]{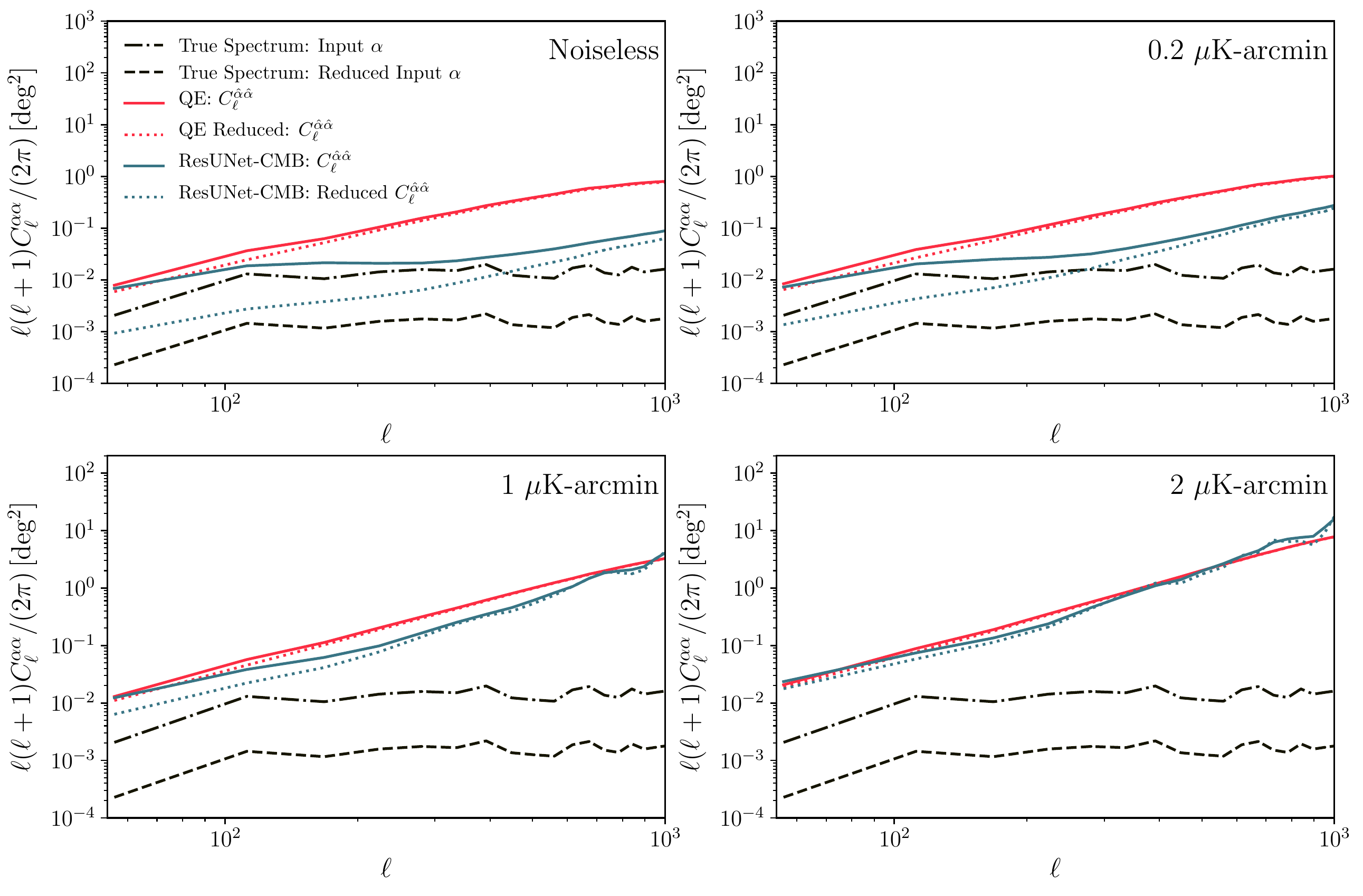}
    \caption{
    Average of reconstructed $\alpha$ power spectra produced with ResUNet-CMB (blue) and quadratic estimator (red) computed with \num{7000} simulated CMB maps, plotted for four experimental configurations.  We present results for maps rotated with two realizations of the true $\alpha$, one using the nominal amplitude $\ell^{2} C_\ell^{\alpha\alpha}/(2\pi)$=0.014~deg$^2$ (solid lines) and the other using a reduced amplitdue $\ell^{2} C_\ell^{\alpha\alpha}/(2\pi)$=0.0016~deg$^2$ (dotted).  We also show the power spectra of the true $\alpha$ maps used for the nominal (black dash-dot) and reduced (black dashed) cases.}
    \label{fig:QE_compare}
\end{figure*}

\begin{figure*}[ht!]
    \centering
    \includegraphics[width=\textwidth]{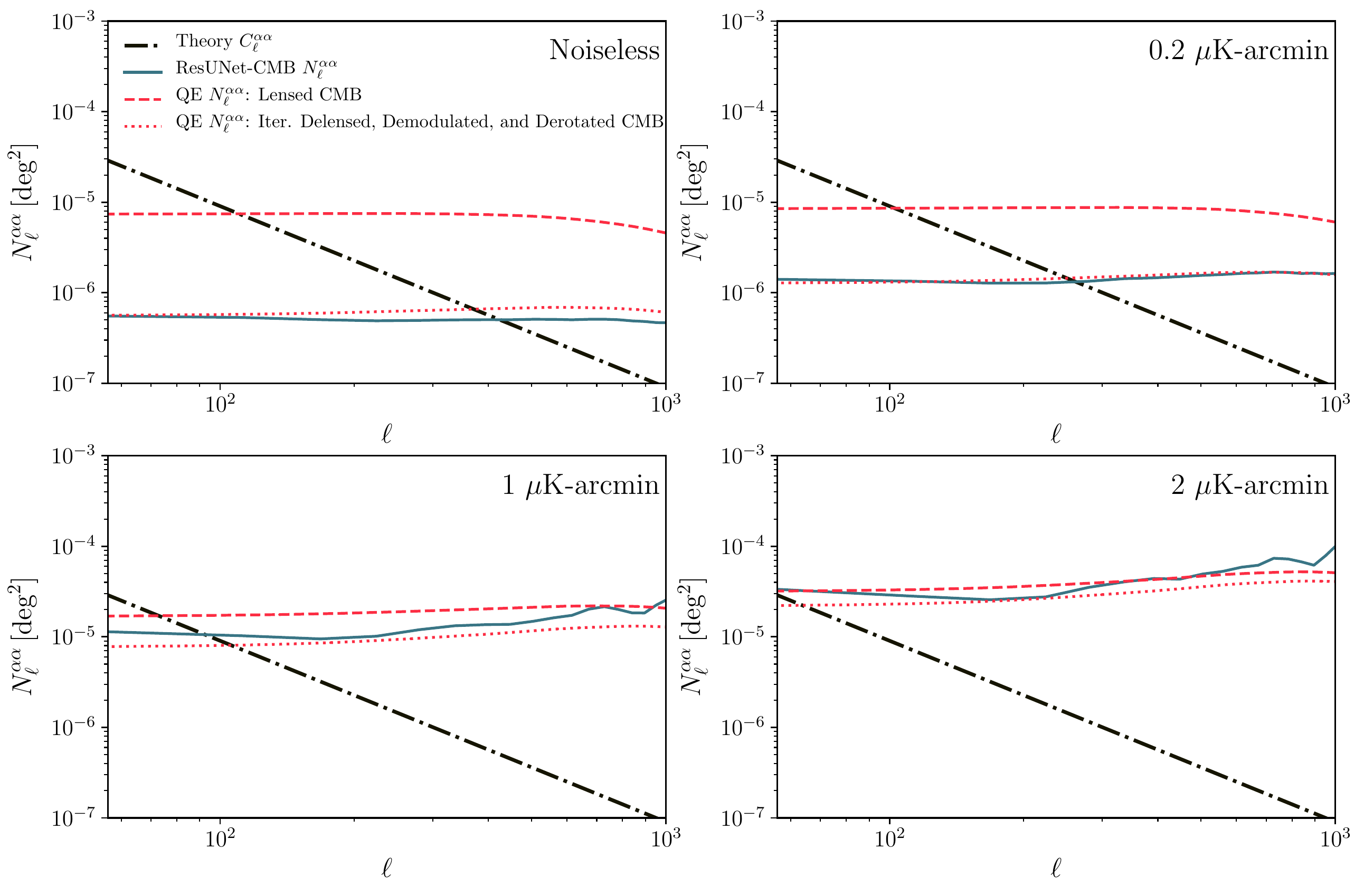}
    \caption{
    Cosmic birefringence reconstruction noise power spectra for ResUNet-CMB (blue); the standard quadratic estimator using lensed, rotated, and modulated CMB spectra (red dashed); and a quadratic estimator using the iterative reconstruction technique described in Sec.~\ref{sec:QE} (red dotted). Also plotted is the theory $\alpha$ spectrum (black dash-dot).}
    \label{fig:nltt_wbias}
\end{figure*}

Finally, we compare the performance of the standard quadratic estimator in reconstructing the $\alpha$ power spectrum to the predictions of ResUNet-CMB.
For the quadratic estimator we use the $EB$ estimator for anisotropic cosmic birefringence, $\hat{\alpha}_{EB}(\vect{\ell})$ (see Eqs.~\eqref{eqn:GeneralQE}, \eqref{eqn:GeneralMinVarEBFilter}, \eqref{eqn:GeneralMinVarEBNorm}, and \eqref{eqn:AlphaModeCoupling}), which is implemented using the software package \texttt{Symlens}\footnote{\url{https://github.com/simonsobs/symlens}}.

As previously mentioned, the direct outputs of ResUNet-CMB exhibit a multiplicative bias.

We must correct this bias in order to make comparisons with the predictions from the quadratic estimator. We follow the same steps that were used for the ResUNet lensing reconstruction by Ref.~\cite{Caldeira:2018ojb} and for patchy reionization in Ref.~\cite{Guzman:2021nfk}, modified appropriately for cosmic polarization rotation.
The biased maps of each distortion field $\theta$ are rescaled by a factor
\begin{equation}
    \label{eqn:map_normalization}
    A_{\ell}^\theta = \left[ \frac{\langle C_{\ell}^{\theta \uthetamap}\rangle}{\langle C_{\ell}^{\theta \theta}\rangle} \right]^{-1} \, ,
\end{equation}

where the $\langle\ldots\rangle$ is an average over the entire prediction data set.

The final unbiased estimate of the anisotropic polarization rotation is given by
\begin{equation}
    \label{eqn:norm_prediction}
    \hat{\alpha}(\vect{\ell}) = A_{\ell}^\alpha \ualphamap(\vect{\ell}) \, .
\end{equation}
The reconstructed $\alpha$ power spectrum predicted by ResUNet-CMB is then
\begin{equation}
    \label{eqn:unbiased_ps}
    C_{\ell}^{\hat{\alpha} \hat{\alpha}} = \left(A_{\ell}^\alpha\right)^{2} (C_{\ell}^{\biasmap{\alpha} \biasmap{\alpha}}) \, .
\end{equation}

In Fig.~\ref{fig:QE_compare} we compare the reconstructed $\alpha$ power spectra produced with the quadratic estimator and with ResUNet-CMB for all four experimental configurations and for two choices of fiducial $C_\ell^{\alpha\alpha}$.
We start by producing a single map of $\alpha$ with a scale invariant power spectrum $\ell^2 C_\ell^{\alpha\alpha}/(2\pi) = 0.014$~deg$^2$.  A second map of $\alpha$ is produced by rescaling the first map such that it has a reduced power spectrum $\ell^2 C_\ell^{\alpha\alpha}/(2\pi) = 0.0016$~deg$^2$ but the same random fluctuations.  A single realization of each of $\kappa$ and $\tau$ are also produced.
Next we generate \num{7000} realizations of the CMB polarization.  Each of these \num{7000} polarization maps is then rotated, modulated, and lensed by the fixed ($\alpha$, $\tau$, $\kappa$) fields and noise is added.  This results in \num{14000} maps of ($\obsmap{Q}$, $\obsmap{U}$), \num{7000} for each amplitude of the $\alpha$ map.
We apply the quadratic estimator to each observed map, the power spectrum of each reconstructed $\alpha$ is computed, and the power spectra are averaged over the \num{7000} CMB realizations for each amplitude of $\alpha$ to produce the red curves shown in Fig.~\ref{fig:QE_compare}.  
The fully trained ResUNet-CMB is applied to the same set of maps, except that a cosine taper is applied before predictions are made. The average of the window-corrected power spectra of the bias-corrected predictions are shown in blue in Fig.~\ref{fig:QE_compare}.  The process is repeated for each of the four experimental configurations we consider.

It is clear from Fig.~\ref{fig:QE_compare} that ResUNet-CMB outperforms the standard quadratic estimator in reconstructing anisotropic cosmic birefringence, especially at low noise.  The power spectra produced from the predictions of the quadratic estimator are almost independent of the true $\alpha$ spectrum, except at low experimental noise and on large scales $\ell<200$.  This is suggestive that the reconstructed power spectra from the quadratic estimator are dominated by noise rather than by signal for the configurations we considered here.  On the other hand, for all but the highest noise experiment we considered, the predictions of ResUNet-CMB are clearly sensitive to the amplitude of the input signal.  The ResUNet-CMB spectra also exhibit lower total (signal plus noise) power across a wide range of scales.  For the 2~$\mu$K-arcmin case, the quadratic estimator and ResUNet-CMB perform nearly identically, with a reconstructed power that does not depend significantly on the signal amplitude, suggesting that the reconstruction is dominated by noise.

We performed a set of tests to check whether $A^\alpha_{\ell}$ depends on the cosmology that is used to train the network.  We trained models for different amplitudes of the birefringence spectrum and compared the value of $A^\alpha_{\ell}$ from each and found them to generally agree, though there is a mild dependence of $A^\alpha_{\ell}$ on the input amplitude of $C_\ell^{\alpha\alpha}$, particularly at the lowest experimental noise levels. We also checked the dependence of $A^\alpha_{\ell}$ on the values of other cosmological parameters.  For significant changes to the assumed cosmology, there were non-trivial changes to $A^\alpha_{\ell}$.  This is to be expected, since the reconstruction noise inevitably depends on the properties of the primordial CMB maps.  This dependence on cosmology appears also in the normalization of the standard quadratic estimator.

When used in a likelihood analysis on real data, one could correct for the cosmology dependence of $A^\alpha_{\ell}$ by accounting for how it depends upon the inputs used to train the network.  A strategy to achieve this is discussed in Ref.~\cite{Caldeira:2018ojb} and briefly summarized here.  First, the network could be trained for a set of cosmologies in the neighborhood of the best fit.  The outputs of this set of trained models could be used to calculate numerical derivatives of the $A^\alpha_{\ell}$ with respect to the fiducial cosmological parameters or input spectra. It would then be possible to apply a correction at every sampled point of the likelihood, as is done with the Planck lensing likelihood, for example (see Appendix~C of Ref.~\cite{Planck:2015mym}). This allows one to avoid biases that may arise from cosmology dependence of $A^\alpha_\ell$ without having to re-train the network for each likelihood sample.

\subsection{Noise}
\label{subsec:Noise}

We now specify the procedure for computing the noise of the ResUNet-CMB reconstruction of cosmic polarization rotation and compare to the reconstruction noise obtained with the quadratic estimator.

The $\alpha$ reconstruction noise for ResUNet-CMB is defined in terms of the bias-corrected power spectrum as
\begin{equation}
    \label{eqn:noise_ps}
    N_{\ell}^{\alpha \alpha} =  \langle C_{\ell}^{\hat{\alpha} \hat{\alpha}} \rangle - \langle C_{\ell}^{\alpha \alpha} \rangle \, .
\end{equation}
where the angle brackets again represent an average over the prediction data set.
Using Eq.~\eqref{eqn:noise_ps}, we calculate the reconstruction noise power spectrum obtained with ResUNet-CMB for each experiment, 
and we show the results in Fig.~\ref{fig:nltt_wbias}. 

We also show in Fig.~\ref{fig:nltt_wbias} two estimates of the reconstruction noise using the quadratic estimator. One estimate, labeled ``Lensed CMB" uses the observed CMB spectra without delensing, demodulation, or derotation.  In this case, the lensing-induced $B$ modes act a source of noise for the sake of the birefringence reconstruction, and $N_\ell^{\alpha\alpha}$ does not decrease much with decreasing noise.  For our choice of parameters, patchy reionization has only a very small effect on the reconstruction noise computed here.  The other case, labeled "Iter. Delensed, Demodulated, and Derotated CMB" uses the iterative reconstruction procedure sketched in Sec.~\ref{sec:QE}.  Decreasing the experimental noise gives significant improvements in the $\alpha$ reconstruction noise obtained with the iterative estimator.  Both versions of the quadratic estimator noise are computed using $\ell_\mathrm{max}=6100$ in order to match the range of angular scales used in the simulations.

For the noiseless and 0.2~$\mu$K-arcmin experiments the ResUNet-CMB reconstruction noise follows closely the noise curve for the iterative reconstruction across a wide range of $\ell$ values. At higher noise levels, the ResUNet-CMB reconstruction noise approaches that obtained from the idealized iterative procedure over a smaller range of $\ell$, though it still outperforms the standard quadratic estimator with no delensing or derotation over a wide range of scales. These results indicate that the deep learning network is able to successfully disentangle and reconstruct the effects of $\alpha$, $\kappa$, and $\tau$ simultaneously, in a nearly optimal way. 

\subsection{Null test}

\begin{figure}[t!]
    \centering
    \includegraphics[width=\linewidth]{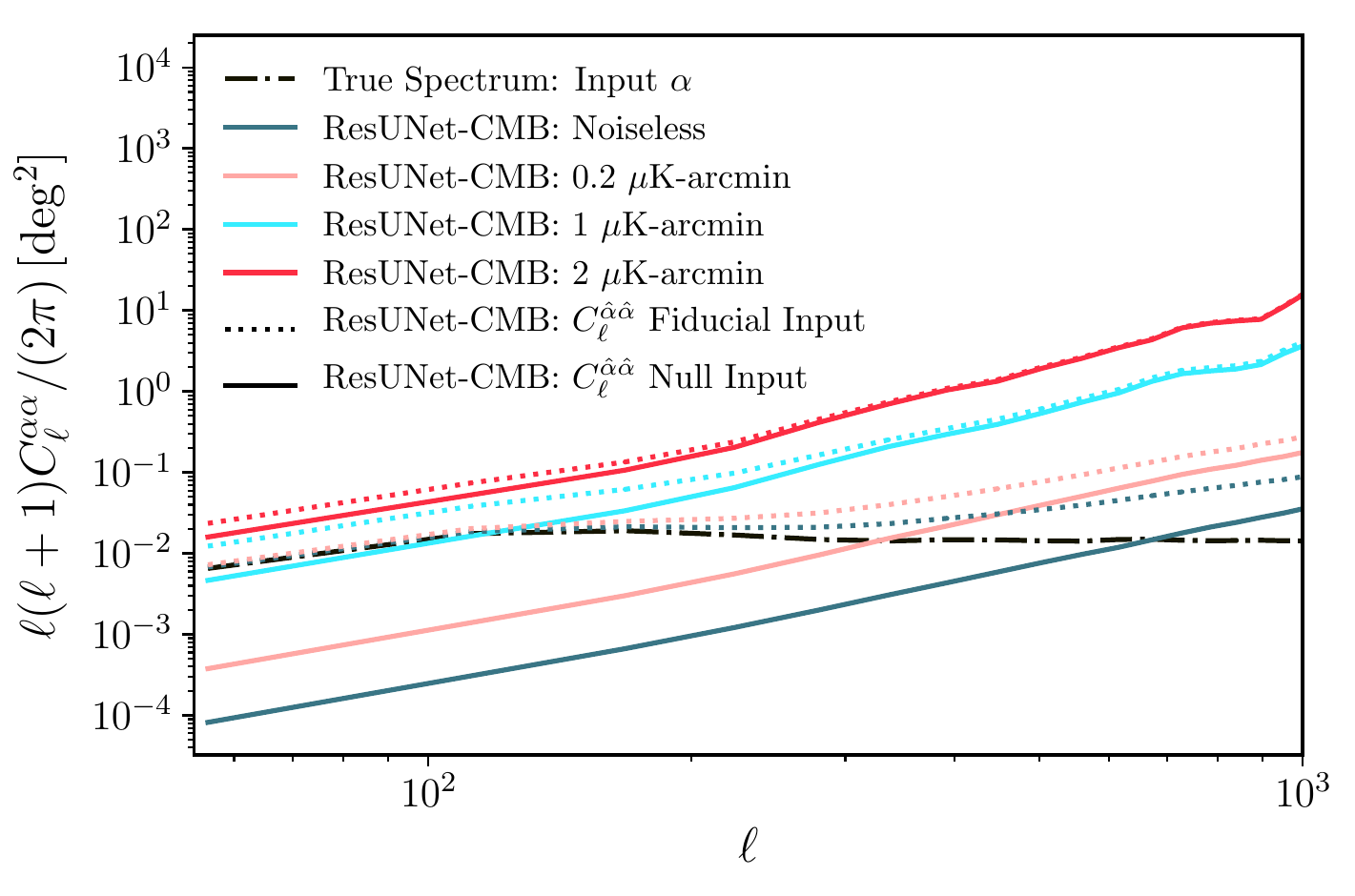}
    \caption{
    Average bias-corrected reconstructed $\alpha$ power spectra of ResUNet-CMB predictions made with two different sets of simulated CMB maps: one set  that includes modulation and lensing but zero rotation (null maps, solid) and on another set that includes rotation, modulation, and lensing (dotted).
    We also plot the true $\alpha$ spectrum used for the rotated maps (black dash-dot).
    }
    \label{fig:null_ps}
\end{figure}

A null test was performed to ensure that the ResUNet-CMB $\alpha$ estimates are due to the presence of a true signal and are not simply an artifact of the network or training process.
We produce \num{7000} simulated CMB polarization maps, ($Q$, $U$), that are modulated by $\tau$ and lensed by $\kappa$ but are not rotated (the $\alpha$ spectrum is set to zero).
We add noise to these maps and use ResUNet-CMB to make predictions as previously described. 
The power spectra of the bias-corrected $\alpha$ predictions are averaged and the results for each experiment are plotted in Fig.~\ref{fig:null_ps}.
For comparison, we also plot the bias-corrected reconstructed power spectra predicted from maps with non-zero $\alpha$ (i.e. the bias-corrected version of the spectra shown in Fig.~\ref{fig:alpha_ps}).

The reconstructed $\alpha$ power spectra from the null maps are consistent with the reconstruction noise spectra shown in Fig.~\ref{fig:nltt_wbias}.
The bias-corrected power spectra predicted from maps with non-zero $\alpha$ are consistent with signal plus noise.
We therefore see that ResUNet-CMB is performing as desired, and the network is recovering the true cosmic polarization rotation signal, rather than reproducing artifacts from the training process.

\begin{figure*}[t!]
    \centering
    \includegraphics[width=\textwidth]{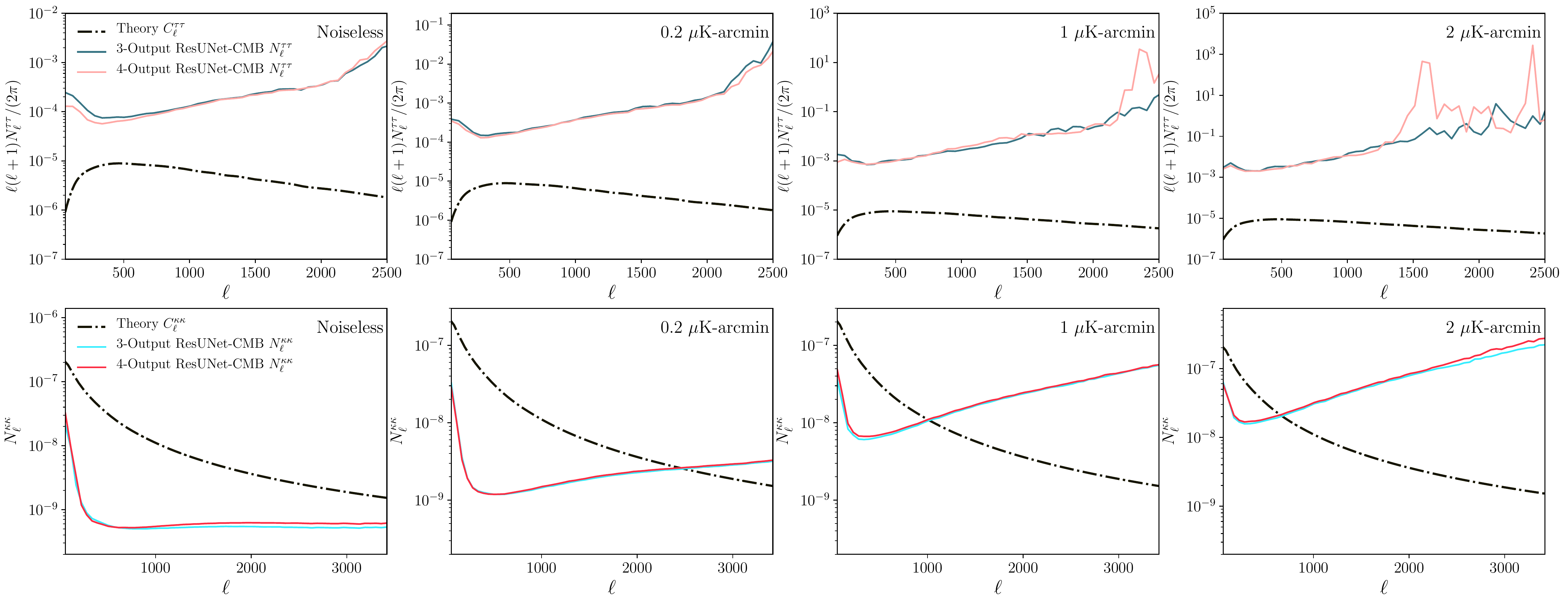}
    \caption{
    Reconstruction noise power spectra of patchy reionization $\tau$ (top row) and lensing convergence $\kappa$ (bottom row) for ResUNet-CMB 3-output (blue, cyan) and 4-output (coral, red) network architectures for each of the four experimental configurations we considered along with the fiducial theory spectra (black dash-dot).
    }
    \label{fig:nltt_k_t_compare}
\end{figure*}

\subsection{Comparison with the 3-output network}
\label{subsec:3to4}
In this section we compare the reconstruction quality of the 3- and 4-output ResUNet-CMB networks by examining the reconstruction noise for the predictions they have in common, $\tau$ and $\kappa$. In Fig.~\ref{fig:nltt_k_t_compare} we show the reconstruction noise as calculated with Eq.~\eqref{eqn:noise_ps}, modified appropriately for $\tau$ and $\kappa$. The top row of figures features the $N_{\ell}^{\tau \tau}$ spectrum while the bottom row contains $N_{\ell}^{\kappa \kappa}$.

Generally, the reconstruction noise for the $\tau$ and $\kappa$ predictions made by the 3- and 4-output networks are in very close agreement.
The predictions of the 0.2~$\mu$K-arcmin experiment were particularly stable, with nearly identical reconstruction noise for the two networks over the entire $\ell$ range we consider. 
The noiseless experiment exhibited a modest increase in $\kappa$ reconstruction noise for $\ell>500$ with the addition of the $\alpha$ output branch.

The small scale $\tau$ predictions for the 1~$\mu$K-arcmin and 2~$\mu$K-arcmin experiments seemed to be most prominently affected by the addition of cosmic birefringence reconstruction.  The 3-output network had been observed to perform worse than the quadratic estimator for these experiments on small angular scales, and the 4-output network seems to exhibit this breakdown at lower $\ell$ values.  However, the $\tau$ reconstruction noise of the 3- and 4-output networks is essentially the same for $\ell<1000$.

Figure~\ref{fig:nltt_k_t_compare} demonstrates that the $\tau$ and $\kappa$ reconstruction have similar variance in the 3- and 4-output networks over a wide range of angular scales. Despite adding another source of statistical anisotropy to the polarization maps and adding an output branch to the network to incorporate its reconstruction, the predictions of $\tau$ and $\kappa$ are mostly unaffected.

\section{Conclusion}
\label{sec:Conclusion}
We extended the ResUNet-CMB architecture from Ref.~\cite{Guzman:2021nfk} by adding an additional output branch to the network for the reconstruction of anisotropic cosmic polarization rotation, $\alpha(\nhat)$. We showed that the ResUNet-CMB network is capable of $\alpha$ reconstruction from modulated, lensed, and rotated CMB polarization maps
with a lower variance than what is achievable with the standard quadratic estimator over a wide range of angular scales.
We also found the ResUNet-CMB estimates of $\alpha$ have a variance that approaches that of an idealized iterative reconstruction method.
The ResUNet-CMB network is able to simultaneously reconstruct three sources of statistical anisotropy, ($\alpha$, $\kappa$, $\tau$), while out-performing the standard quadratic estimator for each output.

The success of ResUNet-CMB with reconstruction of cosmic polarization rotation demonstrates the flexibility and generalizability of the architecture.
We showed how the network faithfully reconstructs the $\alpha$ signal even as we vary the amplitude of the true $\alpha$ applied to the maps.
Unlike gravitational lensing, for which the power spectrum is mostly determined by cosmological parameters that are already well-measured, cosmic polarization rotation results from physics beyond the standard cosmological model, and so we only have observational upper bounds on its amplitude.
It is therefore encouraging that ResUNet-CMB produces results consistent with noise when no cosmic birefringence is present in the maps, implying that it would be a viable tool to search for a first detection of anisotropic cosmic polarization rotation.
Furthermore, the patchy reionization signal that we focused on in Ref.~\cite{Guzman:2021nfk} can be treated as a small perturbation to the lensed CMB, since only a statistical detection would be possible even for a noiseless experiment.  On the other hand, the cosmic birefringence we study here produces much larger effects on the CMB, allowing for a high signal-to-noise reconstruction of the rotation field on large angular scales.
The flexibility of ResUNet-CMB is reinforced by the fact that the network is successful at reconstructing both patchy reionization and cosmic birefringence despite the differing size of their effects for the models we used.

In order to convert the 3-output model of Ref.~\cite{Guzman:2021nfk} to the 4-output model studied here, the only modification to the ResUNet-CMB architecture  was the addition of three convolution blocks that form an extra branch for the $\alpha$ map output.
The data pipeline was altered to include the effects of cosmic birefringence, but no other major changes were required. 
The results show the  variance on the reconstruction of $\alpha$ is smaller than that of the quadratic estimator, and the variances of the $\tau$ and $\kappa$ predictions in the 4-output network are very similar to those of the 3-output network where cosmic polarization rotation was absent.

The success of ResUNet-CMB applied to the new task of cosmic birefringence reconstruction suggests that additional inputs and outputs can straightforwardly be added to the network. 
Experience here shows that these additions need not drastically increase VRAM requirements.
Additional inputs and outputs would enable ResUNet-CMB to be applied to searches for other  sources of secondary CMB anisotropies, including those that are most easily observed by cross-correlating CMB maps with other cosmological surveys.

One limitation of ResUNet-CMB is that it produces point estimates of reconstructed maps with uncertainties that are somewhat opaque. 
As a result, constraining cosmological parameters using the outputs of ResUNet-CMB requires some additional care. 
One path forward would be to incorporate ResUNet-CMB into a Bayesian framework capable of quantifying uncertainties like the one in Ref.~\cite{Millea:2020cpw}. 
Alternatively, ResUNet-CMB could be used as the core component to construct a Bayesian neural network~\cite{gal2016dropout} that is capable of producing a probabilistic distribution of outputs.
Bayesian neural networks have a growing presence in cosmology with applications such as parameter and uncertainty estimation with the CMB~\cite{Hortua:2019ryu, He2018AnalysisOC} and with strong gravitational lensing~\cite{PerreaultLevasseur:2017ltk, Morningstar:2018ase}.
Incorporating ResUNet-CMB into a Bayesian neural network would allow for predictions with well-characterized posterior probabilities.

Building on the success of Refs.~\cite{Caldeira:2018ojb} and \cite{Guzman:2021nfk}, we showed that the ResUNet-CMB architecture can be straightforwardly modified to reconstruct anisotropic cosmic polarization rotation with better performance than the quadratic estimator.
The results presented here give strong motivation to pursue additional applications of ResUNet-CMB.

\vspace*{2em}

\section*{Acknowledgments}
This work is supported by the US Department of Energy Office of Science under grant no.~DE-SC0010129. EG is supported by a Southern Methodist University Computational Science and Engineering fellowship. Computations were carried out
on ManeFrame II, a shared high-performance computing cluster at Southern Methodist University. 

\bibliographystyle{utphys}
\bibliography{bibliography}

\end{document}